\documentclass[preprint2]{aastex}
\usepackage{times}
\slugcomment{Submitted to ApJ}

\shorttitle{R-Process in \Alfven Wave-Driven Winds}
\shortauthors{Suzuki and Nagataki}

\def\mbf#1{\mbox{\boldmath ${#1}$}}
\def\Alfven{Alfv\'{e}n~}

\begin{document}

\title{\Alfven Wave-Driven Proto-Neutron Star Winds And R-Process 
Nucleosynthesis}
\author{Takeru K. Suzuki\altaffilmark{1} and Shigehiro Nagataki\altaffilmark{2}}

\altaffiltext{1}{
Department of Physics, Kyoto University,
Oiwake-cho Kitashirakawa Sakyo-ku, Kyoto 606-8502, Japan; 
JSPS Research Fellow
}
\email{stakeru@scphys.kyoto-u.ac.jp}
\altaffiltext{2}{
Yukawa Institute for Theoretical Physics, Kyoto University,
Oiwake-cho Kitashirakawa Sakyo-ku, Kyoto 606-8502, Japan
}

\begin{abstract}
We propose magnetic proto-neutron star (PNS) winds driven by \Alfven waves 
as well as the neutrino heating as an appropriate site for the r-process 
nucleosynthesis. \Alfven waves excited by surface motions of a PNS 
propagate outwardly, and they heat and accelerate the wind 
by dissipation. 
Compared with the wind purely driven by the neutrino heating, 
larger entropy per baryon and shorter dynamical time scale are achieved, 
which favors the r-process. 
We study reasonable cases that a wave amplitude is 10\% of \Alfven speed 
at the surface to find that 
a PNS with surface field strength, $\gtrsim 5\times 10^{14}$G, 
gives suitable wind properties for the r-process, provided that 
dissipation length of the wave is at most $\sim$ 10 times of 
the PNS radius. 
We also compare properties of transcritical and subcritical winds in 
light of the r-process. 
We finally discuss possibilities of detections of $\gamma$-rays from 
radioactive nuclei and absorption lines due to Ba in supernova remnants which 
possess magnetars. 

\end{abstract}

\section{Introduction}

An origin of elements produced via rapid neutron capture process (r-process) 
still remains a mystery. It is one of the most important astrophysical 
problems to be solved. 
First, most massive nuclei (up to mass number, $A\simeq$ 250) in the 
universe are 
synthesized through the r-process. It is not only scientificly interesting 
but also important to explore how, where, and when they are synthesized. 
Second, some r-process nuclei can be used as chronometers 
\citep{cowan99,wanajo03}. For example,
the half-lives of $\rm ^{232}Th$ and $\rm ^{238}U$ are 1.405$\times 10^{10}$
yr and 4.468$\times 10^{9}$ yr, respectively. 
We can estimate the ages of metal-poor astrophysical objects by their 
observed abundance ratio provided that the mass-spectrum of these produced 
r-process nuclei is predicted precisely. 
Third, chemical evolution of the r-process elements gives us 
a clue to reveal the history of the evolution of our
Galaxy itself ~\citep{ishimaru99,tsy99, argast04}. 

The conditions for the successful r-process nucleosynthesis 
are (Hoffman, Woosley, \& Qian 1997; hereafter HWQ97): 
(i) high density, 
(ii) high entropy per baryon, (iii) short dynamical
timescale, and (iv) small electron fraction, $Y_e$. 
This is because r-process nuclei are synthesized through
the non-equilibrium process of the rapid neutron capture on seed
nuclei, products 
as a consequence of the $\alpha$-rich freezeout (HWQ97).
In other words, an explosive, and neutron--rich site with high entropy 
is a promising candidate for the r-process nucleosynthesis. 

The most probable 
sites, which possibly meet the above conditions, are collapse-driven 
supernovae (SNe; e.g., Woosley et al. 1994) 
and/or neutron star (NS) mergers~\citep{freiburghaus99,rosswog00}.  
However, the collapse-driven SNe are thought to
be more probable sites, 
because metal poor stars have already contained the r-process nuclei.
Mcwilliam et al. (1995) reported that Eu, one of the r-process elements, 
is detected in 11 out of 33 metal-poor stars, 
which proves that r-process nuclei have been already produced
at the very early stage in our Galaxy. Comparing the event rate of
collapse-driven SNe ($10^{-2} \; \rm yr^{-1} \; Gal^{-1}$;
van den Bergh \& Tammann 1991) with that of the NS merger 
($10^{-5} \; \rm yr^{-1}\; Gal^{-1}$; 
van den Heuvel \& Lorimer 1996; Bethe \& Brown 1998),
we can see that collapse-driven SNe are favored since they can supply 
the r-process nuclei at early epochs. 
Moreover, Cowan et al (1999) reported
that the abundance ratio of r-process nuclei in metal poor stars
are very similar to that in the solar system. This indicates that the r-process
nuclei are synthesized through the similar conditions;  
most of the r-process nuclei are from a single 
candidate. So we assume in this paper that most of the r-process nuclei
are synthesized in the collapse-driven SNe.

Many works have been done to uncover the nature of the r-process 
nucleosynthesis in the collapse-driven SNe so far. 
However, 
it seems quite difficult to produce the r-process elements compatible 
with the observations. 
Calculations of the neutrino-driven winds under the Newtonian gravity 
show that entropy per baryon is too small to succeed in the r-process 
(Takahashi et al.1994; Qian \& Woosley 1996; hereafter QW96). 
In order to solve this difficulty, QW96 further included
a post-Newtonian correction to the gravitational force. 
Cardall and Fuller (1997) extended QW96 by considering a 
fully general relativistic treatment. Both showed that the relativistic 
effects favored the r-process nucleosynthesis. 
After an extensive survey of the general relativistic effects, 
Otsuki et al. (2000) gave a quantitative conclusion 
that the r-process was realized in the strong neutrino-driven winds 
($L_{\nu} \sim 10^{52}$erg $\rm s^{-1}$) as long as a massive 
($\sim$2.0 $M_{\odot}$) and compact ($\sim$ 10 km) proto-NS (PNS) is formed.
This is finally confirmed by \citet{sso00}, \citet{wanajo01}, and 
\citet{tbm01}.
Although it is very intriguing as the solution is firstly discovered by the 
introduction of the relativistic effects, 
the above conditions can be satisfied only by 
a soft equation of state (EOS) based on a few nonstandard models for 
sufficiently cold nuclear matter~\citep{wiringa88}.
In fact, the r-process nuclei can not be
produced in the numerical simulations with a normal
EOS~\citep{sso00}. Thus it seems that the difficulty can not be solved
completely by the effects of the general relativity only.

There is only one report of the successful nucleosynthesis 
\citep{woosley94}.
In their numerical simulation, the entropy per baryon becomes high enough to 
produce the r-process nuclei successfully at very late phase of 
the neutrino-driven wind ($\sim 10$ s after the core-collapse). 
However, 
such high entropy 
is inconsistent with a reliable 
analytical formulation based on the Newtonian gravity (QW96).
The general relativistic effects, taken into account in \citet{woosley94}, 
cannot explain the discrepancy, either \citep{otsuki00}. 
Furthermore, nuclei with $A \sim 90$ are overproduced, 
which is incompatible with the solar system abundances. 
The successful mass-spectrum
at the late phase 
would also be destroyed 
by neutral-current neutrino spallations of nucleons from $\rm ^{4}He$ 
\citep{meyer95}. 
Although the result of \citet{woosley94} is remarkable, it would not be 
concluded that the problem is settled completely. 

Facing such difficulties, it is worth exploring 
additional effects that encourage the r-process nucleosynthesis. 
It is argued that the wind properties are improved to be suitable for the 
r-process by a small increase of energy input at a few tens kilometers above 
the surface, which may be operated by wave mechanism and/or magnetic 
activities (QW96; Thompson et al.2001).  
Recent observations show that some fraction of NSs has 
relatively strong magnetic field, $B \gtrsim 10^{14}$G (e.g. Guseinov, 
Yazgan, \& Ankay 2003). It is expected to effect dynamics and energetics 
of the neutrino-driven winds because the magnetic pressure becomes 
comparable to or even larger than the thermal pressure (Thompson 2003). 
As a result, abundances of the r-process products would be different in 
the winds from such magnetic PNSs. 
Let us introduce previous works dealing with the effects of magnetic field 
around the PNSs. 
Nagataki (2001) pointed out importance of the magnetic field
and investigated a steady, subsonic, and rigidly rotating jet,  
although the r-process conditions can not be satisfied 
in the winds. 
Nagataki and Kohri (2001) further examined a steady and subsonic
wind by using magneto-rotational wind solutions (Weber \& Davis 1967), 
although they only explored field strengths, 
$5 \times 10^{11}$ G, which is not dynamically important because of the 
complicated critical point topologies encountered in the steady state 
solution. As a result, they found that the wind properties are still 
inappropriate for the r-process nucleosynthesis. 
Thompson (2003) 
considered an ordered dipole with larger field strength up to $10^{16}$G 
and pointed out an importance of trapping of the plasma heated by the neutrino.  
He estimated enhancement of the entropy due to the trapping to conclude  
that the ordered dipoles with surface fields of $> 6 \times 10^{14}$G
give the suitable circumstances for the r-process nucleosynthesis. 

We propose another mechanism operated by magnetic fields. 
\Alfven waves are supposed to play a role in heating and accelerating  
astrophysical plasma. 
In particular those generated by surface convective motions  
are reliable sources in the solar corona and wind.  
As the waves have momentum and energy, their dissipation leads to 
acceleration and heating of the surrounding plasma 
(e.g. Belcher 1971; Suzuki 2004).
Excitation of \Alfven waves is also introduced in NSs with strong 
magnetic field \citep{td96}.
Then, those \Alfven waves will contribute 
to the heating and acceleration in the magnetosphere, similarly 
to the solar corona. 
To date, however, a role of these waves in PNS winds has not been studied. 
Then, we investigate effects of the \Alfven waves on the physical conditions
such as the entropy per baryon and the dynamical timescale, and 
we discuss whether they can help the synthesis of r-process nuclei.

In section~\ref{model}, we explain the formulation for \Alfven wave-driven 
PNS winds in the
hot bubble. In section~\ref{result}, we show the results.
Discussions and Conclusion are presented in section~\ref{discussions}
and~\ref{conclusion}.

\section{Methods}\label{model}
\subsection{Formulations}
We present formula for the PNS winds in existence of the \Alfven 
waves which excited at the surface. 
Although this is not addressed in this paper, excitation of \Alfven 
waves is studied in NSs with strong magnetic field by several authors 
\citep{td96,tlk02,ko04}.   
A certain fraction of magnetic energy in the core of a NS is 
transferred to seismic waves in the crust. 
Such waves will be coupled with \Alfven waves in the magnetosphere 
\citep{bbgm89}. 
In other words, \Alfven waves would be excited from around the 
surface.  
Moreover, the PNS is convectively unstable\citep{eps79}. 
Drived turbulent-like motions \citep{kjm96} will further contribute to the 
excitation of the \Alfven waves.

We consider one-dimensional magnetic flux tube which is radially 
open. \Alfven waves excited with amplitude of magnetic filed, 
$\delta B_0$, at the PNS 
surface propagate outwardly along with this field line.
The plasma also flows along it when the ideal MHD condition 
is satisfied. 
Then, it can be assumed that all the physical variables depend only on 
radial distance, $r$. 
We also assume steady state with continuous injection of the \Alfven waves 
from the surface. 
We do not consider effects of the rotation of a PNS and the general 
relativity to focus on a role of \Alfven waves in this paper. 

Conservation of magnetic flux, $\mbf{\nabla \cdot B}=0$, gives 
\begin{equation}
\label{eq:mgcns}
B r^2 = B_0 r_0^2 ={\rm const}, 
\end{equation}
where $B$ is radial field strength and $B_0$ is field strength at the PNS 
surface, $r = r_0$. 
Mass conservation under the steady state condition becomes 
\begin{equation}
\label{eq:mscns}
4\pi r^2 \rho v ={\rm const},     
\end{equation}
where $\rho$ is density and $v$ is velocity. 
Momentum conservation is expressed as  
\begin{equation}
\label{eq:mmeq}
v\frac{dv}{dr} = -\frac{GM}{r^2} - \frac{1}{\rho}\frac{dP}{dr}
- \frac{1}{\rho}\frac{dP_w}{dr}, 
\end{equation}
where $M$ is mass of the PNS and $P$ is total pressure of nonrelativistic 
matter, relativistic particles, and photon radiation (QW96). 
$P_w$ is wave pressure which can be written as a function of wave amplitude 
in magnetic filed, $\delta B$, as  $P_w = \frac{\delta B^2}{16\pi}$ 
\citep{lc99}. 
An equation for internal energy, $\cal{E}$, is  
\begin{equation}
\label{eq:engeq}
\dot{q_{\nu}} + \dot{q_{w}} = v(\frac{d \cal{E}}{dr} 
-\frac{P}{\rho^2}\frac{d\rho}{dr} ), 
\end{equation}
where $\dot{q_{\nu}}$ denotes neutrino heating and $\dot{q_{w}}$ is 
heating due to dissipation of the \Alfven waves.

In this study, we consider three processes with respect to 
neutrino heating and cooling, that is, neutrino scattering processes
on electrons and positrons, neutrino absorption on free nucleons,
and electron and positron capture on free nucleons. 
$\dot{q}_{\nu}$ can be calculated as a function of neutrino luminosity, 
$L_{\nu}$, and energy, $\epsilon_{\nu}$, following 
\citet{tubbs75}, \citet{goodman87}, and QW96.
We do not consider heating from neutrino/anti-neutrino annihilation 
and cooling from electron/position annihilation which are taken into 
account in QW96, because they are not effective as shown by Otsuki 
et al. (2000).
We set $\dot{q_\nu}$=0 for
$T \le $0.5 MeV, because free nucleons are bound into $\alpha$-particles
and heavier nuclei and electron-positron pairs annihilate into photons.

The wave heating, $\dot{q}_w (> 0)$, satisfies an equation describing 
variation of wave energy: 
\begin{equation}
\label{eq:wvhtmd}
-\dot{\rho q_{w}} = \mbf{\nabla \cdot F_w}  
-v\frac{dP_w}{dr} \equiv \frac{v_{\rm A}}{v_{\rm A} + v} 
\mbf{\nabla \cdot H_{w}},  
\end{equation}
where $v_{\rm A}=B/\sqrt{4\pi \rho}$ is \Alfven speed and $\mbf{F_w} 
=\frac{\delta B^2}{8\pi}(\mbf{v_{\rm A}} + \frac{3}{2}\mbf{v})$ is 
wave energy flux \citep{jaq77}.
\mbf{H_w} is wave action constant which is expresses as \citep{jaq77}  
\begin{equation}
\mbf{H_w} \equiv \frac{\delta B^2}{8\pi}\frac{(v_{\rm A}+v)(\mbf{v_{\rm A}+v})}
{v_{\rm A}},
\end{equation}
which is a conserved quantity in moving media if the wave does not 
dissipate. 
Physical meaning of eq.(\ref{eq:wvhtmd}) is quite similar to the 1st law 
of thermodynamics; variation of wave energy flux, $\mbf{\nabla\cdot F_w}$, 
is determined by work done by wave, $-v\frac{dP_w}{dr}$, and wave 
dissipation, $-\rho \dot{q}_w$. 
\Alfven wave pressure gradient is written as a function of $\dot{q_w}$ as 
\citep{lc99}
\begin{equation}
\label{eq:wvprs}
\frac{1}{\rho}\frac{d P_w}{dr}= - \frac{\dot{q_w}}{2(v + v_{\rm A})} 
+ \frac{\delta B^2}{32 \pi \rho} \frac{3v + v_{\rm A}}{v + v_{\rm A}}
\frac{1}{\rho} \frac{d\rho}{dr}
\end{equation}

\Alfven waves are supposed to hardly dissipate due to the uncompressional 
character, if they travel in uniform media. 
However, amplitude of those propagating in the stratified plasma, such as 
atmosphere of stellar objects, is amplified so that non-linear damping 
processes becomes important.  
Ingoing \Alfven waves are 
excited by parametric decay instability, 
coupling between the outgoing waves and density fluctuations \citep{gol78}.
These ingoing \Alfven waves interact with the pre-existing outgoing waves to 
lead to turbulent-like cascade and subsequent wave dissipation.   
The parametric decay also generates sound waves, which easily dissipate 
due to their compressive character.  
Linearly polarized waves dissipate through formation of 
fast MHD shocks in magnetically dominated plasma as a consequence 
of nonlinear steepening of the wave fronts \citep{hol82,suz04}. 
Transverse gradient of \Alfven speed leads to resistive and viscous 
dissipation due to phase mixing of \Alfven waves 
traveling along neighboring field lines \citep{hp83}.   
Kinetic effects, such as ioncyclotron resonance, might work in the 
dissipation (e.g. Cranmer, Field, \& Kohl 1999).  
Observation in the solar corona shows an evidence of dissipation 
of \Alfven waves in a region closed to the surface \citep{dtb99}.  
Therefore, \Alfven waves, if they exist, are also expected to dissipate in the 
PNS winds, though the dissipation mechanisms are quite complicated as 
discussed above. 
In this paper, we adopt a simple parameterization for the wave damping  
by using a dissipation length, $l$, 
\begin{equation}
\label{eq:evsw}
H_w = \frac{r_0^2}{r^2}H_{w,0} \exp \left(\frac{r_0 - r}{l}\right),
\end{equation}
where $H_{w,0}$ is wave action at the PNS surface, $r=r_0$.
Equations (\ref{eq:wvhtmd}) and (\ref{eq:evsw}) give an explicit form of 
heating function as 
\begin{equation}
\dot{q}_w = \frac{1}{\rho} \frac{v_{\rm A}}{v + v_{\rm A}} H_{w,0} 
\frac{r_0^2}{r^2}\frac{\exp(\frac{r_0-r}{l})}{l}
\end{equation}
Once initial wave energy flux, $F_{w,0} (\simeq H_{w,0}\;\; {\rm for}\;\; 
v\ll v_{\rm A})$, is fixed, $l$ controls distribution of energy and 
momentum transferred from the waves; heating and acceleration occurs 
in inner regions for cases with smaller $l$, and vice versa.

Let us give a rough estimate of $l$ based on analogy to the nonlinear 
damping of \Alfven waves in the solar corona.  
Generally, $l$ is expected to have a linear dependence on wave length, 
$\lambda = v_{\rm A} \tau$, where $\tau$ is wave period: 
$$ 
l =  f \lambda, 
$$
where $f$ is a proportional coefficient which is expected to  
have a negative dependence on $\delta B/B$ for the nonlinear 
damping. 
\citet{suz04} considered \Alfven wave with $\lambda \sim 10^5$km 
($v_{\rm A}\sim 1000$km/s and $\tau\sim 100$s) as a typical one 
in the solar corona.  
He studied dissipation by fast shocks, one of the nonlinear 
mechanisms, to find that $l$ becomes an order of $R_{\odot}$ for 
$\delta B/B \simeq 0.2-0.3$, 
where $R_{\odot} = 7\times 10^5$km is the solar radius.  
This shows $f \sim 10$, which means that the \Alfven waves 
dissipate typically by 10 times of $\lambda$.  
In PNS winds, timescale for convections can be regarded as one of the typical 
wave periods. Hydrodynamical simulations show that convective cells with 
several kilometers move with velocity $10^3 - 10^4$km/s (Keil et al.1996), 
which gives $\tau \sim 1$ms. 
We consider $B_0\gtrsim 10^{14}$G for \Alfven wave-driven winds, 
which corresponds to $v_{\rm A}\sim 10^4$km/s. Then, typical wave length 
becomes $\lambda = v_{\rm A}\tau \sim 10$km$(=r_0)$. 
If the amplitude is similar to that in the solar wind, the similar 
$f(\sim 10)$ may arise.  
It follows that we can assume $l \sim 10 r_0$ as a fiducial value.   
However, $l$ would vary a lot if waves with different $\tau$ 
dominantly worked. 
Therefore, we consider three cases, $l=(5,10,30) \times r_0$ in this 
paper.  


\subsection{Numerical Integration}
We describe a practical method to determine the wind structures 
for various inputs of \Alfven waves. 
Adopted parameters for a PNS and the input neutrinos are 
summarized in tab.1. 
We set the density at the inner boundary to be $10^{10}$ g/cm$^3$,
following the result of Wilson's numerical simulation in Woosley et
al. (1994) (see also Otsuki et al. 2000). 
Temperature at the surface is derived from energy balance between the neutrino 
heating and cooling, following QW96:
$$
T_{0} = 1.19 \times 10^{10} 
\left[ 1 + 
\frac{L_{\nu_{e}}}{L_{\bar{\nu}_{e}}}
\left( \frac{\epsilon_{\nu_{e}, \rm MeV}}{\epsilon_{\bar{\nu}_{e}, \rm MeV}}
\right)^2
\right]^{\frac{1}{6}} 
$$
\begin{eqnarray}
L^{\frac{1}{6}}_{\bar{\nu}_e, \rm 51}
R^{-\frac{1}{3}}_{\rm 6} \epsilon^{\frac{1}{3}}_{\bar{\nu}_e, \rm MeV} \;\;\;
\rm \left [  K     \right ],
\label{eqn16}
\end{eqnarray}
where $L_{\nu, \rm 51}$ is the individual neutrino luminosity in $10^{51}$
$\rm ergs \; s^{-1}$, $R_{6}$ is the neutron star radius in $10^{6}$ cm 
($=10$km),
$\Delta$ = 1.293 MeV is the neutron-proton mass difference,
and $\epsilon_{\nu, \rm MeV}$ is a neutrino energy in MeV. We assume that
the neutron star radius is equal to the neutrino sphere radius. 

Amplitude of \Alfven waves is set to be $\delta B_0/B_0 = 0.1$ at 
the surface, 
although it is quite uncertain.   
$\delta B_0/B_0 = 0.1$ indicates that velocity amplitude is 10\% of 
the phase speed (\Alfven speed in this case), which is often seen in 
astrophysical objects; for example, at the solar surface speed of 
the convective motions ($\sim 1$km/s) is $\sim 10$\% of the sound speed 
($\sim 10$km/s).  
By the assumption of the constant $\delta B_0/B_0$ the input wave energy flux, 
$F_{w,0}\simeq H_{\rm w,0}$, is proportional to $B_0^3$ (note that 
$v_{\rm A,0} \propto B_0$).

Integration starts from the surface for an initial guess of velocity, $v_0$. 
Temperature is integrated according to eq.(\ref{eq:engeq}). 
Background magnetic field, $B$, is fixed by the conservation of magnetic 
flux (eq.(\ref{eq:mgcns})). 
$\delta B$ is determined in order to satisfy the dissipation model 
of \Alfven wave (eq.(\ref{eq:evsw})). 
We integrate velocity by an equation, transformed 
from the momentum equation (\ref{eq:mmeq}) with help of eq.(\ref{eq:wvprs}), 
$$
\frac{1}{v}\frac{dv}{dr} = \left[\frac{1}{r}(2 v_{\rm s}^2 + 2 w^2
-\frac{GM}{r} ) - \frac{\dot{q}_{\nu} + \dot{q}_{w}}{3v} 
+ \frac{\dot{q}_{w}}{2(v + v_{\rm A})}\right]  
$$
\begin{equation}
\label{eq:trdv}
\left[v^2 - v_s^2 - w^2\right]^{-1},
\end{equation}
where $v_{\rm s} = (\frac{4}{3}\frac{P}{\rho})^{1/2}$ is adiabatic sound speed 
and 
$$w \equiv \sqrt{\frac{\delta B^2}{32 \pi \rho} 
\frac{3v + v_{\rm A}}{v + v_{\rm A}}}.$$
is a variable in unit of velocity, indicating contribution of the \Alfven waves. 
When deriving eq.(\ref{eq:trdv}) from eq.(\ref{eq:mmeq}), we have assumed 
that pressure by relativistic electrons and positrons and photon radiation 
dominates the other components, and the degeneracy is negligible, namely, 
\begin{equation}
\label{eq:apprs}
P\simeq \frac{11}{4}\frac{a T^4}{3}, 
\end{equation}
where $a$($= 7.56\times 10^{-15}$erg cm$^{-3}$K$^{-4}$) is 
a Stefan Boltzmann radiation constant. 
The above assumption is applicable only for $T\gtrsim 0.5$MeV 
because otherwise the electron-positron pairs annihilate and the remained 
electrons become nonrelativistic. Although the effect of the pair 
annihilation gives 
a modification of eq.(\ref{eq:apprs}) by a factor of $(11/4)^{1/3}$,  
this seems to influence on the dynamics quite little (\cite{sso00}). 
Contribution of the non-relativistic gas to the pressure is also small 
for $T\gtrsim 0.1$MeV in our \Alfven wave-driven winds.
This is because density in the \Alfven wave-driven 
winds is smaller than the neutrino driven-winds on account of rapid 
acceleration.
Once $v$ is determined by eq.(\ref{eq:trdv}), $\rho$ is 
derived from the mass conservation equation (\ref{eq:mscns}).

\begin{table}[]
\caption{Adopted Parameters. $r_0$ is PNS radius, $\rho_0$ is surface density, 
and $M$ is PNS mass in unit of solar mass, $M_{\odot}$. 
$L_{\nu}$ is neutrino luminosity per flavor and $\epsilon_{\nu}$ is neutrino 
energy. }
\begin{tabular}{c|c}
\hline
$r_0$ & 10km\\
$\rho_0$ & $10^{10}$g cm$^{-3}$\\
$M$ & $1.4 M_{\odot}$ \\
$L_\nu$ & $10^{51}$erg s$^{-1}$\\
$\epsilon_{\nu_e}$ & 12MeV \\
$\epsilon_{\bar{\nu}_e}$ & 22MeV \\
$\epsilon_{\nu_\nu},\epsilon_{\bar{\nu}_\nu},\epsilon_{\nu_\tau},
\epsilon_{\bar{\nu}_\tau}$ & 34MeV \\
\hline
\end{tabular}
\label{tab}
\end{table}

For the velocity structure we both consider transcritical and subcritical solutions. 
When deriving transcritical solutions, 
a critical point is derived from the condition that both numerator and 
denominator of eq.(\ref{eq:trdv}) are zero. 
Then, we iteratively determine an 
initial guess of $v_0$ to smoothly pass through the critical point 
by a shooting method.
On the other hand, subcritical solutions cannot be uniquely determined. 
We derive subcritical solutions for given mass flux by reducing $v_0$ 
from the transcritical values.

We have checked an accuracy of the integration by monitoring whether 
how precisely a Bernoulli equation shown below is fulfilled 
as a function of $r$:
\begin{equation}
\label{eq:bernol}
\left[\frac{v^2}{2} + \frac{T S}{m_{\rm N}} - \frac{G M}{r} 
+ \frac{F_W}{\rho v}\right]_{r_0}^{r} = \int_{r_0}^{r} dr 
\frac{\dot{q}_{\nu}}{v} ,  
\end{equation}
where $m_{\rm N}$ is the nucleon rest mass and 
$S=m_{\rm N}\frac{4P}{\rho}$ is entropy per baryon 
(for non-degenerate and relativistic gas).
We have found that our numerical integration satisfies the above equation 
at least within 3\% which is acceptable for the purpose of our model 
calculations.

\subsection{Condition for Entropy}
When we discuss whether r-process nucleosynthesis can occur or not, we 
use a criterion given by HWQ97, which is written as
\begin{equation}
\label{eq:crtrp}
S \gtrsim 2\times 10^3 Y_e \left(\frac{t_{\rm exp}}{s}\right)^{1/3}
\end{equation}  
for $Y_e \ge$0.38, where $S$ is measured in unit of 
Boltzmann constant, $k_{\rm B}$, and $t_{\rm exp}$ 
is expansion time scale defined as time elapsed from $T=9\times 10^9$K 
to $T=2.5 \times 10^9$K.
This is the condition for production of the r-process nuclei
with mass number $A \sim$200 (QW96).

Electron fraction $Y_e$ at the time when
$\alpha$-rich freezeout takes place can be estimated as (QW96)
\begin{eqnarray}
Y_{e} = \left( 1 + \frac{L_{\bar{\nu}_{e}}}{L_{\nu_{e}}}
\frac{\epsilon_{\bar{\nu}_{e}, \rm MeV} - 2\Delta + 
1.2 \Delta^2/ \epsilon_{\bar{\nu}_{e}, \rm MeV}  }{\epsilon_{\nu_{e}, \rm MeV}
+ 2\Delta + 1.2 \Delta^2/\epsilon_{\nu_{e}, \rm MeV}}  \right)^{-1} .
\label{eqn14}
\end{eqnarray}
When we substitute the parameters shown in tab.1, we estimate $Y_e$ = 0.43, 
at which the r-process criterion (eq.(\ref{eq:crtrp})) is applicable. 
In realistic situations, 
$L_\nu$ and $\epsilon_\nu$ may vary, hence, we consider 
cases with $Y_e = 0.4 -0.5$

\section{Results}\label{result}
In this section, we examine how the dissipation of the \Alfven waves 
modifies the wind structures. In particular we 
discuss possibilities of the r-process nucleosynthesis in light of 
the criterion, eq.(\ref{eq:crtrp}), for $S$ and $t_{\rm exp}$. 

The transcritical solutions are generally stable if 
sufficient energy injection is given and the winds blow into free space. 
In the PNS winds, however, ejected shells exist in the outer region 
and interaction between the shells and winds leads to formation of 
termination shock.  
We expect the flow stays transcritical if a position of the shock 
is far outside of the critical point. 
Otherwise, the shock works as an outer boundary so that the flow might 
become subcritical eventually. 
Thus, we firstly study properties of transcritical flows in detail 
(\S\ref{sc:trscr}), and then, show different aspects of subcritical 
flows for comparison (\S\ref{sc:subcr}). 

\subsection{Transcritical Flows}
\label{sc:trscr}
\subsubsection{Effects of \Alfven Waves}
\label{sc:gnrs} 
Figure 1 illustrates how the wind properties change by the effects 
of \Alfven waves. 
As for \Alfven wave, we consider two cases of $B_0 = 3\times 10^{14}$G 
(dotted) and $B_0 = 5\times 10^{14}$G (solid).
In both case the same dissipation length, $l=10r_0$, is adopted. 
Wind purely driven by the neutrino heating is also shown for 
comparison (dashed). 
Since $F_{w,0}\propto B_0^3$, the input \Alfven wave energy flux of 
case with $B_0=5\times 10^{14}$ is larger by a factor of 5 than that of 
case with $B_0=3\times 10^{14}$.
Wind structures ($\rho$, $T$, \& $v$), entropy per baryon ($S$), and 
heating ($\dot{q}_{\nu}$ \& $\dot{q}_w$) are compared in top, middle, 
and bottom panels, respectively. 

One sees in bottom panel that the wave heating is distributed around 
$\sim l (=10 r_0)$, in contrast to the neutrino heating which is 
localized close to the surface. 
Then, the wind structures shown in top panel are modified in 
relatively outer region by the \Alfven waves.  
A clear difference is seen in velocity 
structure. The \Alfven waves directly accelerate the materials by 
wave pressure gradient and the faster winds are attained as input energy 
of \Alfven wave 
increases. 
Accordingly, 
$t_{\rm exp}=8.9\times 10^{-3}$s in the case with larger input of 
\Alfven waves is 
much shorter than $t_{\rm exp}=4.4\times 10^{-2}$s in the case without 
the waves. 
This indicates that the wave dissipation realizes non-equilibrium 
circumstances for the nuclear reactions, which encourages
the r-process nucleosynthesis. 
Density shows more rapid decrease by the presence of the \Alfven waves to 
satisfy the mass conservation (eq.(\ref{eq:mscns})). 
Temperature is also slightly 
lower because adiabatic cooling effectively works due to the rapid 
expansion (i.e. acceleration) of the plasma.


Middle panel shows that the \Alfven waves give a great influence 
on $S$ in the wind. 
First, $S$ is no more a constant even in the outer region where the neutrino 
heating is inefficient because 
the wave heating still occurs there. 
Second, $S$ is larger except in a region closed to 
the surface for larger injection of \Alfven waves. 
This trend needs to be interpreted in detail because it is opposed to 
that anticipated from the neutrino heating (QW96) . 
Variation of entropy is given by 
\begin{equation}
\label{eq:entvar}
\frac{dS}{dr} = \frac{m_{\rm N}\dot{q}_{\nu}}{v T} 
+ \frac{m_{\rm N}\dot{q}_w}{v T}, 
\end{equation}  
which is derived from the energy equation (\ref{eq:engeq}). 
For the neutrino heating, larger $\dot{q}_{\nu}$ leads to 
larger $v$ and $T$ in the winds. As a result, 
$\frac{\dot{q}_{\nu}}{v T}$ itself becomes  
smaller for larger $\dot{q}_{\nu}$ and final $S$ is also smaller (QW96). 
On the other hand in the wave heating, an increase of $v$ as a consequence of 
an increase of $\dot{q}_w$ is smaller than that of $\dot{q}_w$ itself 
as described below. 

Based on our wave heating model, eq.(\ref{eq:wvhtmd}), $\dot{q}_w$ is 
proportional to $B_0^3$ for the assumed constant initial amplitude 
$\delta B_0/B_0 = 0.1$. 
As for $v$, a rough dimensional estimation can be done in a following way. 
Without \Alfven waves, velocity, $v_1$, at an arbitrary position $r=r_1$ can 
be derived from the Bernouilli equation (\ref{eq:bernol}) as 
\begin{equation}
\label{eq:berpr1}
\frac{v_1^2}{2} + \frac{T_1 S_1}{m_{\rm N}} - \frac{G M}{r_1}
= -\frac{G M}{r_0} + \int_{r_0}^{r_1}dr \frac{\dot{q}_{\nu} }{v}, 
\end{equation}
where subscript '1' denotes physical variables at $r_1$ and we 
have used the fact that the gravitational potential term dominates at the 
surface, $r=r_0$. 
Let us assume that velocity increase to $v_1 + \Delta v_1$ by an input of \Alfven waves with energy flux, $F_{w,0}$, at $r_0$. 
Then, 
$$
\frac{(v_1+\Delta v_1)^2}{2} + \frac{(T_1 + \Delta T_1) (S_1 + \Delta S_1)}
{m_{\rm N}} - \frac{G M}{r_1} + \frac{F_{w,1}}{\rho_1 v_1 }
$$
\begin{equation}
\label{eq:berpr2}
= 
-\frac{G M}{r_0} + \int_{r_0}^{r_1}dr 
\frac{\dot{q}_{\nu} }{v'} + \frac{F_{w,0}}{\rho_0 v_0 }, 
\end{equation}  
where $\Delta T_1$ and $\Delta S_1$ are modifications of $T_1$ and $S_1$, 
and we use $v'$ instead of $v$ in the integral of the neutrino term 
because the velocity changes from that in the pure neutrino-driven wind 
(eq.(\ref{eq:berpr1})). 

The wave energy converts to both kinetic energy (first term on the left of 
eq.\ref{eq:berpr2}) and enthalpy  
(second term) of the flow. We assume that a fraction, $\epsilon$, is 
transferred to the kinetic part, whereas $\epsilon$ is automatically 
calculated in the numerical integration to satisfy the wave energy 
equation (\ref{eq:wvhtmd}).   
Comparing eqs.(\ref{eq:berpr1}) and (\ref{eq:berpr2}), we have 
\begin{equation}
\label{eq:difvel}
\frac{(v_1 + \Delta v_1)^2}{2} - \frac{v_1^2}{2} = \epsilon 
(\frac{F_{w,0}}{\rho_0 v_0} - \frac{F_{w,1}}{\rho_1 v_1}), 
\end{equation}
where we neglect the difference of the velocity appearing in the 
neutrino term. 
If $\Delta v_1 \ll v_1$, $\frac{\dot{q}_w}{v}$ in eq.(\ref{eq:entvar}) 
simply proportional to $\dot{q}_w \propto B_0^3$, hence, 
an increase of $B_0$ raises $S$. 
If $\Delta v_1 \gg v_1$, equation (\ref{eq:difvel}) gives 
$\Delta v_1 \propto \sqrt{F_{w,0}} \propto B_0^{3/2}$ for constant $v_0$, 
provided that $\frac{F_{w,0}}{\rho_0 v_0} \gg \frac{F_{w,1}}{\rho_1 v_1}$. 
In the real situation $v_0$ has a positive correlation with $B_0$ so that 
the dependence of $\Delta v_1$ on $B_0$ is weaker. 
Therefore,  $\frac{\dot{q}_w}{v}$ in eq.(\ref{eq:entvar}) is at least in 
proportion to $B_0^{3/2}$, showing that larger field strength directly 
leads to an increase of $S$. In the \Alfven wave-driven winds, temperature 
is also slightly lower due to the larger adiabatic cooling (fig.1). 
This effect further contributes to the increase of $S$ as expected from 
eq.(\ref{eq:entvar}). 
In conclusion, \Alfven waves effectively raise entropy in the winds 
to give the suitable circumstances for the r-process.    


Since $S$ varies in a region where the r-process occurs, it is not 
straightforwards whether the r-process criterion, eq.(\ref{eq:crtrp}), 
is directly applicable. In this paper, we adopt $S$ at $T=0.2$MeV as a 
necessary condition, based on the fact that a ratio of neutrons to seeds, 
a key to determine the final abundance of the r-process elements, 
is once fixed through the $\alpha$-rich freeze-out in 
$T\gtrsim 0.2$MeV. 
The figure exhibits that $S(T=0.2{\rm MeV})$ (filled triangles) increases 
for larger input of the \Alfven waves. 
In particular, case with $B_0 = 5\times 10^{14}$G gives  
$S = 170$, which  
satisfies the criterion, eq.(\ref{eq:crtrp}), 
for the r-process nucleosynthesis, if $Y_e \le 0.41$. 

\subsubsection{Dependence on Dissipation Length}
\label{sc:dslg}
Figure 2 compares the wind properties for different dissipation lengths, 
$l=5$ \& 30 but the same field strength, $B_0 = 5\times 10^{14}$G. 
As illustrated in bottom panel, $l$ controls the distribution of heating; 
the waves dissipate more efficiently around $\sim l$ so that the heating 
and acceleration (not shown) are most effective there.    
For smaller $l$, the \Alfven waves influence on the wind in an inner part; 
higher $v$, smaller $\rho$, and $T$ are obtained as seen in top 
panel, which is consistent with the trends obtained when considering 
larger wave heating (\S\ref{sc:gnrs}). $S$ is also larger in the inner region 
(middle panel) as the wave heating is enhanced there.  
However, these trends become reverse in the outer region as the wave dissipation becomes 
efficient for larger $l$ case. 

The wind properties in $T\gtrsim 0.2$MeV are important with respect 
to the r-process nucleosynthesis, since the neutron-to-seeds 
ratio is determined there.   
Wind structure is greatly improved there in case with $l=5$ to give 
smaller $t_{\rm exp}$ and larger $S$ for the 
successful nucleosynthesis, while the modification is smaller in case 
with $l=30$ as the wave dissipation occurs in farther outside. 
Therefore, rapid wave dissipation (short $l$) is favored for the 
r-process.

\subsubsection{$B_0$ \& $l$ for R-process}
In fig.3 we show result of the wind properties for various 
$B_0$ and $l$ in $t_{\rm exp}$-$S(T=0.2{\rm MeV})$ plane,  
overlayed with the r-process criterion 
(eq.(\ref{eq:crtrp}))
for $Y_e=0.4$ and $0.5$. 
We do not show results with $B_0>6\times 10^{14}$G, because both 
wind velocity and \Alfven speed are approaching to the light speed, $c$, 
and our non-relativistic treatment is not valid. (The final flow speed becomes 
$\simeq (0.4-0.5)c$ and the \Alfven speed at $r\sim 5r_0$ becomes 
$\simeq c$ for $B_0=6\times 10^{14}$G.)  
The figure shows that the \Alfven waves play a role for 
$B_0 \gtrsim 10^{14}$G.  
Larger $B_0$ and smaller $l$ lead to larger $S(T=0.2{\rm MeV})$ and 
smaller $t_{\rm exp}$ as described in \S\ref{sc:gnrs} and \S\ref{sc:dslg}. 
Models with $B_0 \gtrsim 5\times 10^{14}$G and 
$l\lesssim 10 r_0$ satisfies the r-process condition, 
eq.(\ref{eq:crtrp}),  
for $Y_e = 0.4$. 
The condition for $Y_e = 0.5$ is also satisfied with slightly larger 
$B_0$ $(= 6\times 10^{14}$G) if $l\lesssim 10 r_0$, as dependence on 
$B_0$ is sensitive.

The value, $B_0 \simeq 5\times 10^{14}$G, can be understood by a simple 
energetics consideration. 
It is useful to introduce a plasma $\beta$ value, $\beta 
\equiv P/(B^2/8\pi)$, a ratio of plasma pressure to magnetic pressure. 
Let us consider $\beta$ at $r = 10 r_0$: 
\begin{equation} 
\label{eq:betad}
(\beta_{10})^{-1} 
\sim 10 \left(\frac{B_0}{5\times 10^{14}{\rm G}}\right)^2 
\left(\frac{0.15{\rm MeV}}{T_{10}}\right)^4, 
\end{equation}
where subscript '10' denotes quantities measured at $r=10 r_0$. 
Here, we have used the conservation of the magnetic flux (eq.(\ref{eq:mgcns})) 
to relate $B_{10}$ with $B_0$, and we assume that the radiation 
pressure dominates in $P$. 
Equation (\ref{eq:betad}) shows that the flow at $10 r_0$ is 
dominated by the magnetic field energy by 10 times for 
$B_0 = 5\times 10^{14}$G. 
Transfer of small fraction 
of the magnetic energy to the radiation plasma 
gives a great influence on the energetics and dynamics. 
\Alfven waves play a role in this energy transfer in our framework, and 
the wind properties are improved to be suitable for the r-process 
nucleosynthesis. 
Sensitive behaviors of the wind parameters on $B_0$ seen in 
fig.\ref{fig:taudp2} can also be interpreted by squared dependence on 
$B_0$ in eq.(\ref{eq:betad}).  

\subsection{Subcritical Flows}
\label{sc:subcr}
We have discussed the r-process nucleosynthesis in the transcritical 
flows so far.  This is reasonable if the plasma expands into space with 
much lower density. In the real situations, however, 
the PNS winds would be interrupted by the ejected shells existing in the 
outer region to form the termination shocks.  
If a position of the shocks are not far from the critical point, 
the winds might become subcritical flows eventually. 
\citet{tsy02} pointed out that these dynamics at the outer boundary 
gave an influence on the r-process nucleosynthesis. Hence, it is worth 
studying the dynamics of the subcritical flows with \Alfven waves.  
In this subsection, we examine how the structures in the subcritical 
flows are modified compared with the transcritical case and argue 
possible influences on the r-process nucleosynthesis. 

Infinite numbers of subcritical solutions exist for different choices of $v_0$. 
Physically, a certain solution is selected by conditions at the 
outer region, such as pressure in the shell ejected by the SN explosion. 
The shell moves outward and the pressure would decrease as it expands. 
Then, it is required to treat the outer boundary in a time-dependent 
manner (e.g. Sumiyoshi et al. 2000) for accurate calculations.  
However, we construct subcritical wind structures under the steady state 
condition by reducing $v_0$ in an ad hoc way for simplicity's sake. 

Figure 4 compares subcritical flows with the transcritical one 
for $B_0=5\times 10^{14}$G and $l=10 r_0$. 
As for the subcritical flows, we show cases with smaller 
$v_0$ by 0.1\%(case A), 1\%(B), and 2.5\%(C) than that of the transcritical case. 
The figure shows that the wind structures in the inner region ($r\lesssim 
10r_0$) are almost identical, so that $t_{\rm exp}$ differs only by 
a factor of 1.6 even between case C and the transcritical flow. 
Clear difference is seen in the outer region. 
Speed of subcritical flows in the outer region is slower, by definition,
than the transcritical flow. Accordingly, the density is much higher 
to satisfy the mass conservation. Temperature is also higher because 
the wave dissipation works in the plasma heating rather than 
the wind acceleration. 
The difference in the entropy ($\propto T^3/\rho$) is 
small because the change of $\rho$ is compensated by that of $T$, whereas 
the subcritical cases give slightly larger $S(T=0.2{\rm MeV})$. 
The r-process criterion, eq.(\ref{eq:crtrp}), is influenced quite 
little as the tiny variances 
of $S$ and $t_{\rm exp}$ further cancel out. 

We would like to discuss implications on the r-process nucleosynthesis 
inferred from these different wind properties. 
The r-process nuclei are synthesized by following two sequential processes 
\citep{wh92}.   
In the inner region with higher temperature ($T\gtrsim 0.2$MeV), 
$\alpha$-process proceeds to synthesize the seed nuclei with mass 
number, $A\sim 100$. 
In the outer region with lower temperature ($T\lesssim 0.2$MeV), 
reactions concerning 
charged particles cease and the neutron capture by the seed nuclei, 
namely the r-process in the narrow sense, dominantly occurs to produce 
the heavy nuclei. 
Since the dynamics in the inner region resemble each other, 
nucleosynthesis by the $\alpha$-process is expeted to be similar so 
that the neutron-to-seed ratio fixed at $T\simeq 0.2$MeV is not different 
so much whether the flows are transcritical or subcritical.  

A main contrast would arise in the final neutron capture in the 
outer region. 
An advantage of the subcritical flows 
is that the seed nuclei have the sufficient time to capture all the 
ambient neutrons due to the slower flow velocity (Yamada 2004).
Temperature ($T\sim 0.1$MeV) of the subcritical winds is 
also reasonable for the neutron capture \citep{wanajo02}. 
We expect that the subcritical winds might be more favorable for the 
synthesis of the r-process nuclei with $A\sim 200$, whereas detailed 
calculation of nuclear reaction is desired to give the final 
conclusion.

\section{Discussions}
\label{discussions}

\subsection{Configurations of Magnetic Fields}
In this paper, we have only considered flux tubes radially open 
as a simplest case. 
If some portions of the surface are covered with closed magnetic fields, 
the open flux tubes expand super-radially
(e.g. Goldreich \& Julian 1969 for the dipole case). 
We can raise mainly two effects with respect to such super-radial expansions. 
First effect is that it reduces \Alfven speed, leading to shorter dissipation 
length of \Alfven waves. 
It may result in 
larger entropy due to an increase of 
the heating in the inner regions. 
Second effect is that it enhances adiabatic loss to reduce the wind speed
\citep{suz04} and the temperature. 
Then, $t_{\rm exp}$ would be changed, whereas detailed investigation 
is required to tell how it is modified.  

If the plasma in the closed region affects the PNS winds by 
magnetic reconnections, situations are quite different. 
The plasma in the closed regions are once heated by the wave dissipation 
as well as the neutrino heating and their sudden break-up leads to inpulsive 
heating. This may also give suitable sites for the r-process 
nucleosynthesis as pointed out by \citet{tho03}. 

\subsection{PNS Rotation}
Massive stars are known to rotate rapidly~\citep{tassoul78}. 
Recent numerical calculations also show that the iron core is rotating prior 
to the collapse~\citep{heger00,heger04}.
If the rotation is taken into account, 
jet-like explosion possibly takes place~\citep{kotake03,takiwaki04}. 
It is shown that such a jet-like explosion makes the products of explosive 
nucleosynthesis
change so much~\citep{nagataki97,nagataki98,nagataki00,nagataki03}.
Although the r-process is not favored in steady, subcritical, and rigidly 
rotating jets in which a balance is maintained between the centrifugal and 
magnetic forces (Nagataki 2001), 
there will be still a possibility of the successful r-process nucleosynthesis 
in a jet-like explosion. In particular, we should investigate dynamical 
jets without assuming a steady state. 
Numerical simulations with rotation and
magnetic field are highly desired to investigate such a possibility.

There will also be a possibility of the r-process nucleosynthesis in 
the winds on the equatorial plane. 
Nagataki \& Kohri (2001) have investigated only steady and
subcritical wind solutions which use the model of Weber \& Davis (1967), 
although they are not relevant for the r-process. 
Otherwise if we investigated dynamical winds without assuming the 
steady-state,
we may find a situation where successful r-process nucleosynthesis occurs,
as pointed in Nagataki \& Kohri (2001).

\subsection{Possibilities of Gamma-Ray Observation}
Qian, Vogel, \& Wasserburg (1998; hereafter QVW98) discussed 
possibilities of detection of 
$\gamma$-ray emitted by the radioactive decay of the synthesized r-process 
nuclei in SN remnants (SNRs). They estimated $\gamma$-ray emissions of several 
isotopes with half-life, $\tau_{\rm hl}\sim 10$yr, assuming the 
solar r-process abundances are explained by the 
accumulation of the Galactic SNe. The derived values are an order of 
$\sim 10^{-7}\gamma$cm$^{-2}$s$^{-1}$ if a SN explodes at a distance 
of 10kpc, which is below the sensitivity of International Gamma-Ray 
Astrophysics Laboratory (INTEGRAL), but at the 
level of future proposed telescope, ATHENA.   

Our results indicate that magnetic PNSs could produce more r-process 
elements than the usual PNSs with $B_0 \sim 10^{12}$G. 
We would like to give a rough estimate on the $\gamma$-ray emissions 
from these magnetic PNS in the similar way to that adopted in 
QVW98.
Subgroups of the NSs called anomalous X-ray pulsars (AXPs) and soft 
gamma repeaters (SGRs)
are inferred to have strong magnetic field, $\sim 10^{14 - 16}$G
\citep{dt92,vg97,kou98,zh00}. 
Although it is still premature to carry out statistical 
analysis on number fraction of these magnetic NSs, 
a rough estimate can be done. 
Comparison between the nearby AXPs and SNRs gives, $\lesssim 1/50$ 
of SNe have magnetic NSs (Guseinov et al.2003). 
\citet{aro03} adopted more conservative limit for birth rate of the magnetars, 
$10^{-5} - 10^{-3}$yr$^{-1}$, which indicates that 0.03 -- 3\% of the SNe 
remain magnetic PNSs for assumed rate of SN, 0.03yr$^{-1}$.    
 
Let us consider an extreme case that all the r-process elements 
have been synthesized by the magnetic PNSs. If we adopt 1\% for the number 
fraction of the magnetic PNSs as a fiducial value, the required r-process 
yield per PNS for the solar abundance is increased by a factor of 100 against 
the case in which all the PNSs produce the r-process elements equally. 
Therefore, expected intensities of $\gamma$-ray lines from several isotopes 
become $\sim 10^{-5}\gamma$cm$^{-2}$s$^{-1}$ for a distance of 10kpc. 
Two AXPs are detected within $\sim$ 7kpc (Guseinov et al.2003). 
Their field strength is derived as $\simeq 8\times 10^{14}$G (1E 1841-045) 
and $\simeq 7\times 10^{13}$G (1E2259+586) from the 
$P-\dot{P}$ diagram, whereas, these values could be larger if an effect 
of non-uniform rotation is taken into account \citep{lb04}. 
Unfortunately, estimated ages, $> 1000$yr, of the SNRs \citep{vg97,par98} 
are much longer than the half lives ($\sim 10$yr) of the quoted nuclei 
in QVW98, 
so that it is impossible to detect the $\gamma$-ray by INTEGRAL.
However, our estimate implies that $\gamma$-ray in future magnetic PNSs 
associated with SNe in the Galactic disk are detectable by INTEGRAL, 
whereas the possibility of the occurrence is low due to small birth rate 
($\sim 10^{-5} - 10^{-3}$yr$^{-1}$).

QVW98 further considered long lived radioactive nuclei, 
$^{126}$SN with $\tau_{\rm hl} = 1.14 \times 10^5$yr. 
Their advantage is that $\gamma$-ray can be observed for a relatively longer 
period in spite of the weaker emissivities. 
Then, we do not have to wait another SN to explode. 
We can estimate the $\gamma$ ray emission to be
$\sim 2 \times 10^{-8} (\delta M/5 \times 10^{-5} M_{\odot}) (7{\rm
kpc}/d)^2 \gamma$cm$^{-2}$s$^{-1}$ for 1E 1841-045~\citep{sanbonmatsu92},  
and $\sim 5.5 \times 10^{-8} (\delta M/5 \times 10^{-5} M_{\odot}) 
(4{\rm kpc}/d)^2 \gamma$cm$^{-2}$s$^{-1}$ for 1E2259+586~\citep{coe92}, 
where $\delta M$ is the average amount of
mass of a radioactive $r$-process nucleus produced in a magnetar 
(QVW98). 
Thus it will be difficult to detect $\gamma$-ray lines from these long
lived radioactive nuclei by INTEGRAL. However, the $\gamma$ rays from 
these AXPs might be marginally detectable by ATHENA whose sensitivity 
is $\sim 10^{-7}$ $\gamma$cm$^{-2}$s$^{-1}$ at $E_{\gamma} \sim 100-700$ 
keV in the future, whereas a large uncertainty of $\delta M$ should be 
reduced by theoretical efforts. 
 

\subsection{Absorption Features of Supernova Spectrum}

There are reports that absorption lines of Ba[II] and Ba[III] are
detected in SN1987A as well as in other Type II-P supernova
in the early phase of the explosion \citep{mazzali92,mazzali95}, which may
indicate that r-process nucleosynthesis occured in SN1987A. 
\citet{shigeyama01} estimated the amount of synthesized Barium in
SN1987A is about 6$\times 10^{-6}M_{\odot}$. Thus if a SN that
contains a magnetar explodes in nearby galaxies and ejected
more r-process elements, the spectrum should show clearer absorption 
features due to Ba. 
In this case, Sr, $\sim 10$\% of which is produced
by the r-process (note that Sr is mostly produced by the s-process), would
be observable as well \citep{shigeyama01}.


\section{Conclusion}\label{conclusion}
We have investigated possibilities of the r-process nucleosynthesis 
in the PNS winds driven not only by the neutrino heating but also 
by the \Alfven waves. Those excited by the surface motions propagate upwardly 
to heat and accelerate the winds through their dissipation. 
As a result, larger $S$ and smaller $t_{\rm exp}$ can be accomplished. 
We have considered the waves with initial amplitude, $\delta B_0/B_0 =0.1$ 
and construct 1 dimensional spherical symmetric wind structures under 
the steady state condition.  
We have shown that field strength, $B_0 \gtrsim 
5\times 10^{14}$G, satisfies the criterion for the r-process by HWQ97, 
if the \Alfven waves dissipate around a distance, $\lesssim$ 10 times of 
the radius, which is reasonable for the waves with period of 1ms. 
We also compared transcritical and subcritical flows with respect to the r-process. 
We have found that the final neutron captures are affected by the types 
of the flows, while the neutron-to-seed ratio after the $\alpha$-process is 
expected to be similar. 

Our results imply that the magnetic PNS is a more appropriate site for  
the r-process nucleosynthesis than PNSs with usual field strength 
$\sim 10^{12}$G. 
Based on this consideration, we have given an estimation of $\gamma$ ray 
emissions by the radioactive nuclei from AXPs (and SGRs), relevant candidates 
for the magnetic NSs. 
Our estimates show that  
$\gamma$ rays from the short-lived nuclei could be detected in future 
SNe associated with the magnetic PNSs by the present telescope, INTEGRAL, 
if one happens in spite of the low possibility, while those from the 
long-lived nuclei would be marginally detectable in the present AXPs by 
the future telescope, ATHENA.    

We are grateful to S. Yamada, T. Murakami, D. Yonetoku, S. Wanajo, 
T. Kajino, and T. Shigeyama for useful discussions. 
Comments raised by an anonymous referee have helped us quite a lot to  
improve the paper.  
This work is in part supported by a Grant-in-Aid for the 21st Century COE 
``Center for Diversity and Universality in Physics'' at Kyoto University.
T.K.S. is financially supported by the JSPS Research Fellowship for Young
Scientists, grant 4607. S.N. is partially supported by Grants-in-Aid
for the Scientific Research from the Ministry of Education, Science
and Culture of Japan through No.S 14102004, No. 14079202 and
No. 16740134.

\begin{figure}
\figurenum{1} 
\epsscale{0.75} 
\plotone{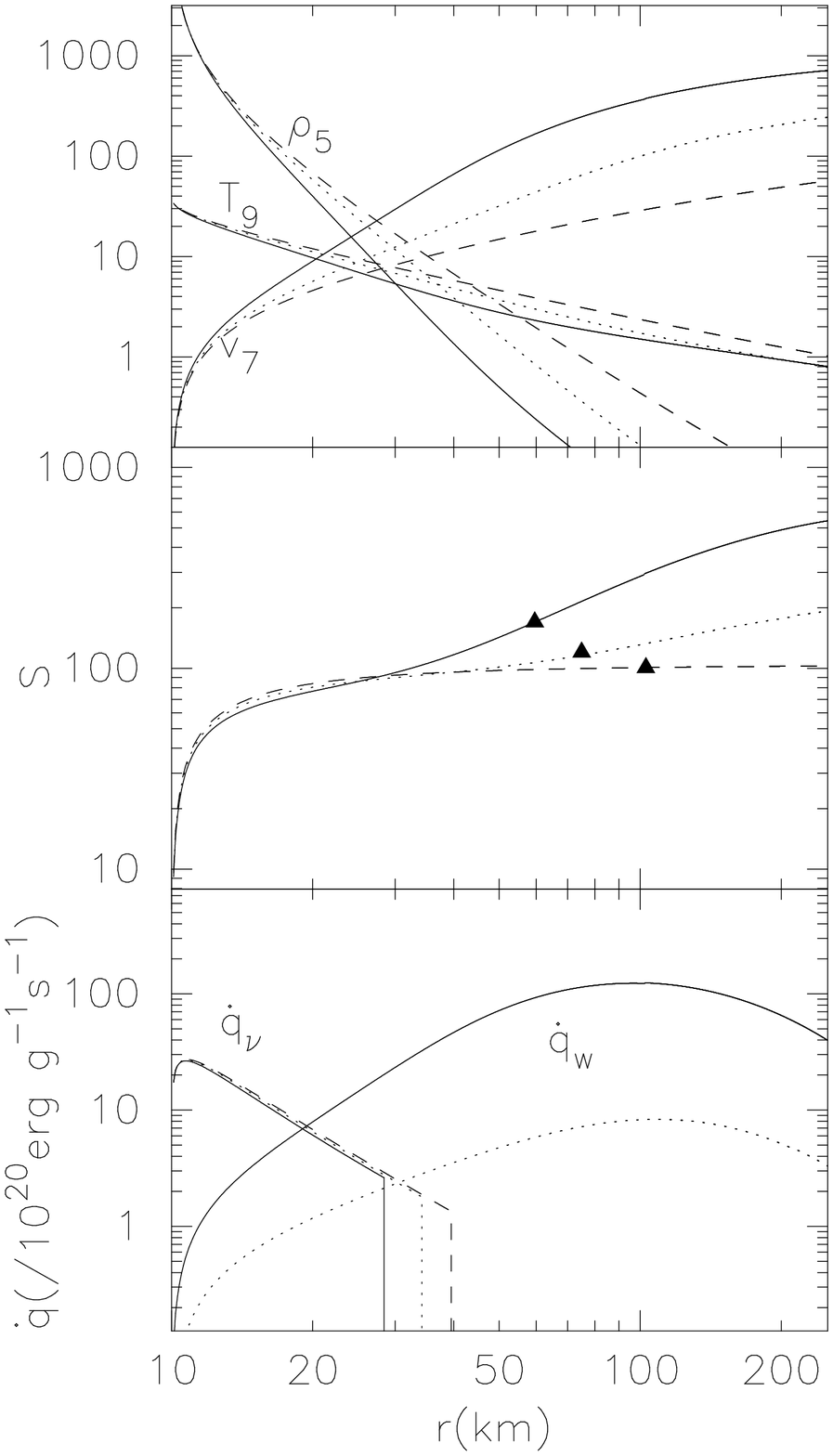}
\caption{Properties of \Alfven wave-driven winds for $B_0 = 
5\times 10^{14}$G (solid) and $3\times 10^{14}$G (dotted). 
Dissipation length is set to be $l=10 r_0$ for both cases. 
Result of neutrino driven-wind (without \Alfven waves) is also 
displayed for comparison (dashed).
{\it top} : Comparison of density 
in $10^5$g cm$^{-3}$, temperature in $10^9$K, and velocity in 
$10^7$cm s$^{-1}$. 
{\it middle} : Comparison of entropy per baryon in unit of $k_{\rm B}$. 
Triangles indicate positions where $T=0.2$MeV. 
{\it bottom} Comparison of heating ($10^{20}$erg/g s) by the neutrino 
($\dot{q}_{\nu}$) and wave dissipation ($\dot{q}_w$).  
}
\label{fig:taudp}
\end{figure}

\begin{figure}
\figurenum{2} 
\epsscale{0.75} 
\plotone{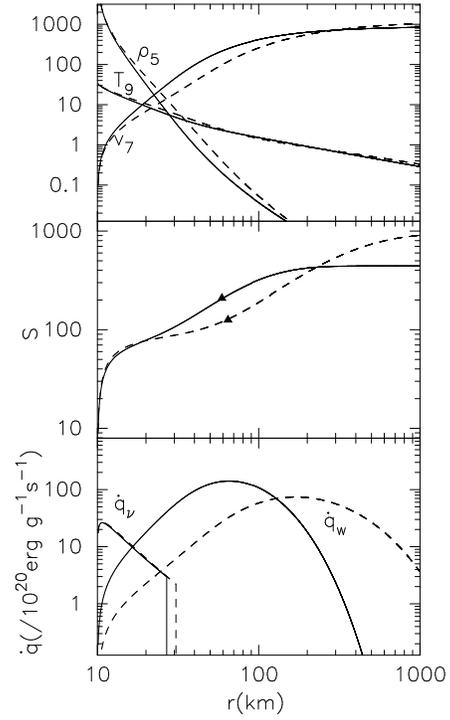}
\caption{Comparison of the wind properties for different 
dissipation lengths, $l=5$ \& $30 \times r_0$ but 
the same $B_0 = 5\times 
10^{14}$G. Each panel is the same as in fig.1 but the horizontal and 
vertical scales are different.  
}
\label{fig:taudp}
\end{figure}

\begin{figure}
\figurenum{3} 
\epsscale{0.8} 
\plotone{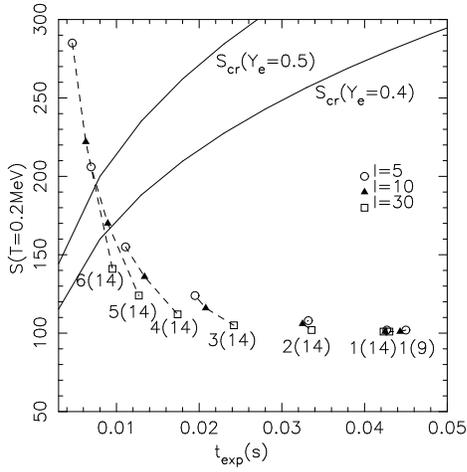}
\caption{$t_{\rm exp}$(s) (X-axis) and $S$ at $T=0.2$MeV (Y-axis) for 
various parameters of \Alfven waves. Open circles, 
filled triangles, and open squares are results with $l=5$, 10, \& 30, 
respectively. Numbers denote magnetic field strength at the surface. For 
example $6(14)$ indicates $B_0=6\times 10^{14}$G. Results with the same $B_0$ 
are connected by dashed lines. Solid lines are the conditions for the 
r-process by HWQ97 (eq.\ref{eq:crtrp}) for $Y_e=0.4$ \& 0.5. 
}
\label{fig:taudp2}
\end{figure}

\begin{figure}
\figurenum{4} 
\epsscale{0.75} 
\plotone{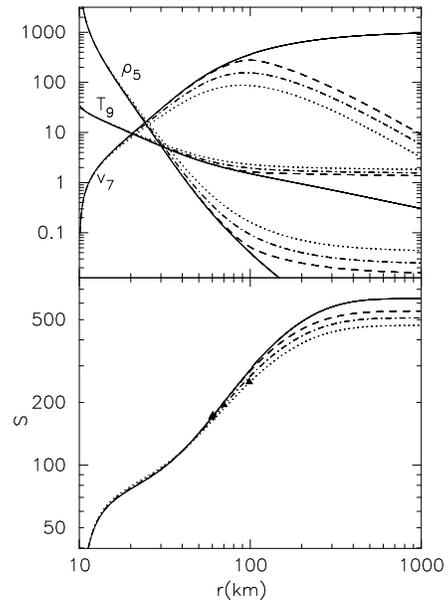}
\caption{Wind properties of subcritical flows 
in comparison with transonic flow (solid) for $B_0 = 5\times 
10^{14}$G and $l=10 r_0$. 
Dashed, dot-dashed, and dotted lines are cases with reducing $v_0$ by 0.1\%, 
1\%, \& 2.5\%, respectively.
Upper and lower panels are the same as in top and middle panels in 
fig.1, but the horizontal and vertical scales are different. 
}
\label{fig:taudp3}
\end{figure}

\end{document}